\documentclass[twocolumn,aps,prb,floatfix,showpacs]{revtex4}

\usepackage{graphicx}

\usepackage{amssymb}

\begin{document}


\title{Effect of Hole Doping on the Electronic Structure of Tl2201}

\author{S. Sahrakorpi,}
\author{Hsin Lin,}
\author{R.S. Markiewicz,}
\author{and A. Bansil}

\address{Physics Department, Northeastern University, Boston MA
02115, USA }

\begin{abstract}

  We discuss doping dependencies of the electronic structure and Fermi
  surface of the monolayer Tl$_{2-x}$Cu$_x$Ba$_2$CuO$_{6+\delta}$
  (Tl2201).  The TlO bands are found to be particularly sensitive to
  doping in that these bands rapidly move to higher energies as holes
  are added into the system. Such doping effects beyond the rigid band
  picture should be taken into account in analyzing and modeling the
  electronic spectra of the cuprates.

\end{abstract}

\date{\today}




\maketitle

The Fermi surface (FS) of Bi$_{2-x}$Pb$_x$Sr$_2$CaCu$_2$O$_{8+\delta}$
(Bi2212) predicted by band theory is well-known to display Bi-related
pockets around the $M(\pi,0)$-point, which have never been observed
experimentally. We have shown recently\cite{Lin2006} that when the
effects of hole doping via Pb/Bi substitution or by adding excess
oxygen are included, the BiO pockets are lifted above the Fermi energy
($E_F$). With decreasing hole doping the BiO bands drop below the
$E_F$ and the system self-dopes below a critical hole concentration in
the sense that further reduction in hole content no longer reduces the
doping of the CuO$_2$ planes. The aforementioned lifting of BiO
pockets is a consequence of substantial Coulombic effects which come
into play due to rearrangement of charges with changing hole
concentration in the system. Ref.~\cite{Lin2006} presents detailed
results on Bi2212 and provides more generally a first-principles route
for exploring doping dependencies of electronic structures of the
cuprates. Here we extend the discussion of Ref.~\cite{Lin2006} to
consider doping effects in monolayer Tl2201.

Insofar as technical details are concerned, we have employed both the
Korringa-Kohn-Rostoker (KKR) and linear augmented plane wave (LAPW)
full potential band structure methodologies where all electrons are
treated self-consistently\cite{Bansil1999,wien2k}. The KKR scheme is
well known to be particularly suited for a first-principles treatment
of the electronic structure of substitutionally disordered
alloys. Effects of hole doping were included within the framework of
the virtual crystal approximation (VCA) which, as discussed in
Ref.~\cite{Lin2006}, is expected to provide a reasonable description
of the cation-derived bands near the $E_F$ in the
cuprates\cite{VCA}. The tetragonal lattice data of
Ref.~\cite{Shimakawa1990} for an overdoped sample of
Tl2201\cite{Shimakawa1990,tetra} was employed in the computations.

Fig.~1(a) shows the familiar band structure of undoped (half-filled,
$x=0.0$) Tl2201. It displays the CuO band characteristic of the
cuprates, which lies at $\approx$~$-$1.3~eV at $\Gamma$, yields the
van Hove singularity (VHS) at $\approx$~$-$0.5~eV around the
$M(\pi,0)$-point, and forms along the $M(\pi,0)-X(\pi,\pi)-\Gamma$
line the non-symmetric inverted parabola. In addition to this CuO
band, a second band, which is TlO-related, is seen to drop below the
$E_F$ at $\Gamma$, giving rise to a $\Gamma$-centered electron pocket,
which has not been observed
experimentally\cite{Hussey2003,Plate2005}. The problem is similar to
that of BiO pockets noted above in connection with Bi2212 except that
here the pocket is centered around the $\Gamma$ rather than the
$M(\pi,0)$-point.

Fig.~1(b) presents results for 24\% hole doped Tl2201, where the
effects of doping have been taken into account. The TlO band has now
moved $\approx$~0.9~eV above the $E_F$ and the TlO pocket of Fig.~1(a)
has disappeared from the electronic structure at an intrinsic
level. At 24\% hole doping, the rigid band shift (lowering) of the
$E_F$ will also empty the pocket\cite{Plate2005}, but the present
calculations show that this pocket will be removed very rapidly with
doping as the TlO band moves to higher energies. Such an effect of
doping beyond the rigid band model is less dramatic on the CuO band,
although the CuO band also undergoes change in shape in that it
narrows slightly with doping.

\begin{figure}
\begin{center}
  \resizebox{8.5cm}{!}{\includegraphics{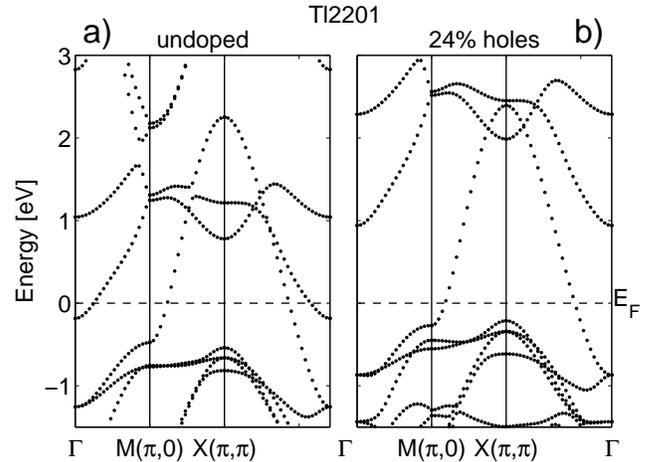}}
\end{center}
\caption{
  \label{fig:1} Band structures of (a) undoped and (b) 24\% hole doped
  Tl2201 along the main symmetry directions in the 2D Brillouin zone
  of the tetragonal lattice.}
\end{figure}
We comment briefly on our theoretical predictions in relation to
relevant experimental results on overdoped Tl2201 single crystals. The
shape and three-dimensionality of the FS derived from the band
structure of Fig.~1(b) is in general accord with that reported in
Ref.~\cite{Hussey2003} via angular magnetoresistance oscillation
(AMRO) measurements, including delicate variations in the shape of the
3D FS half-way between nodal and antinodal directions, as well as with
the angle-resolved photoemission (ARPES) results of
Ref.~\cite{Plate2005}. However, even though the general shape and the
3D nature of the FS is captured by LDA, the theoretical FS is more
squarish than the experimental one (i.e.  closer to $X$ in $M-X$
direction and closer to $\Gamma$ in $\Gamma-X$ direction), possibly
reflecting strong correlation physics beyond the conventional
picture. Notably, the ARPES lineshapes for emission from the $E_F$ in
Tl2201 show relatively little broadening even in the antinodal
region\cite{Plate2005}, indicating that interlayer coupling effects in
Tl2201 are smaller than in La$_{2-x}$Sr$_x$CuO$_4$
(LSCO)\cite{Bansil2005,Sahrakorpi2005}.

In summary, we have shown with the example of Tl2201 that the
electronic structures and Fermi surfaces of the cuprates possess
significant doping dependencies resulting from generic Coulombic
effects that come into play as the electronic system rearranges itself
with changing hole content, and that these effects are particularly
pronounced on the cation derived bands (e.g. the TlO states in the
case of Tl2201). The rigid band model, where one assumes the band
structure to be independent of doping, has usually been the basis for
discussing doping evolution of the electronic spectra of the cuprates
in much of the existing literature. Our study shows that the rigid
band model is fundamentally flawed and that doping effects on the
electronic structure beyond the rigid band model should be accounted
for in analyzing and modeling cuprate physics.

We thank A. Damascelli for a discussion concerning the shape of the
Tl2201 FS. This work is supported by the US Department of Energy
contract DE-AC03-76SF00098, and benefited from the allocation of
supercomputer time at the NERSC and the Northeastern University's
Advanced Scientific Computation Center (ASCC).


\begin{thebibliography}{00}
\bibitem{VCA}
%
Specifically, Tl atoms with nuclear charge of $Z=81$ were replaced by
pseudo-atoms with $Z=80.88$ to account for 24\% hole doping per Cu of
CuO$_2$-plane.
\end{thebibliography}
\end{document}